# Commissioning, Performance, and Effect of the Quench Current-boosting Device on a Dedicated Superconducting Magnet

S. Stoynev, M. Baldini, S. Feher

*Abstract*—Superconducting magnet training is one of the accelerator related issues attracting attention due to significant operational costs and time budget associated to it. It is especially worrisome that magnets based on the "next-generation" Nb$_3$Sn technology are affected by long training. While various efforts are underway to better understand and resolve the problem a parallel path could also be investigated, a path bypassing the issue. Following the concept of fast induced over-current during magnet powering, FNAL has developed an upgradable capacitor-based device to discharge through a superconducting magnet at quench detection or operator chosen time. The 0.4 F/1 kV device has been tested on a 1-m-long dipole-coil in a "mirror" magnet configuration and conclusive results on magnet training elimination have been observed. In this paper we discuss the main characteristics of the device, compare simulated response and actual performance, elaborate on test drivers and outcomes. Next steps and perspectives for future use are debated.

## I. Introduction

**B**UILDING state-of-the-art superconducting accelerator magnets is a delicate process and, among other things, it involves a careful "pre-stress" setting aiming to minimize conductor degradation and optimize performance. This step could be considered "pre-conditioning" [1, Chapter 1.2.5] of the magnet. However, one can expand the meaning of "pre-conditioning" to include any process that would affect magnet performance positively. Here we choose to separate action taken before magnet powering and during magnet powering - "pre-conditioning" and "operational conditioning", respectively. "Pre-conditioning" was considered and experimented with in past [2], [3]. It is known to the authors that "operational conditioning" was conducted in the past too, but we could find no clear reference. That for instance includes manipulating current ramping (levels, rates) to avoid lower current quenches though it is purely an investigative technique.

 The work described in the present paper builds up on [2] where capacitors were discharged through a magnet as a "pre-conditioning" step. We do this as "operational conditioning". In our case, a capacitor is discharged through a magnet being powered at user defined time, for instance at quench detection. This boosts the current through the magnet to levels depending on circuit parameters, including magnet parameters. Such a boost could effectively increase the magnet quench current associated to a given quench although a "quenchless" mode will be debated later too. Given that magnet training is understood as the steady increase of quench current after consecutive spontaneous quenches, changing quench current level is an important lever to affect training. Authors are aware of possible phase delays between operating current and local force [4] in pulsed mode although expectations pointed to time delays of low tens of milli-seconds at most. Our own measurements of magnetic field in magnet bore and magnet current during the sharp decrease of current during system trips and quenches (with immediate magnet protection) did not indicate any phase delays between magnet current and bore field beyond our resolution of a couple of milli-seconds. If present, significant local phase delays, originating from decaying eddy currents, could suppress the Lorentz force peak experienced by the superconductor/coil and diminish the effect of fast current boost.

The work to bring the boost ideas to fruition was supported by LDRD funds at FNAL and the resulting device [5] is in effect a pulsed power supply integrated into the main power supply CPS3 [6] with ability to be controlled independently. We call it Quench Current-boosting Device, or QCD. QCD, has similarities to CLIQ [7] but is a very different device. Apart from being used for different purposes, there are two other major differences: a) the QCD boost current is the same through the whole magnet, i.e. there is no difference in magnet Lorentz force distribution with respect to "regular" ramp-up; b) QCD has no current/voltage oscillation features.

This paper describes the first application of QCD on a dedicated superconducting magnet, points to relevance of simulations before and during testing, reviews choices made during testing and results obtained; those are followed by a discussion on future use of QCD and the technique itself.

## II. QCD Commissioning and Magnet Testing

### A. QCD preparations and simulations

QCD [5] started working as a unit towards the end of 2021 and was gradually integrated to CPS3. After all major components and sub-circuits were verified and tested, the device went through various full circuit examinations, including powering. Before using it on a superconducting magnet, a

This work was supported by Fermi Research Alliance, LLC, under Contract DE-AC02-07CH11359 with the U.S. Department of Energy, Office of Science, Office of High Energy Physics.(*Corresponding author: Stoyan Stoynev*).

 The authors are with Fermilab National Accelerator Laboratory, Batavia, IL 60510 USA (e-mail: stoyan@fnal.gov).

 Digital Object Identifier will be inserted here upon acceptance.

conventional accelerator magnet (fabricated for use in the FNAL accelerator complex) was utilized as a load to demonstrate operation. Engineers from Accelerator Division of FNAL, who developed the device engineering concepts and worked through the process all the way to commissioning, also helped with circuit simulations. LTSpice software [8] was employed allowing to explore sensitivity and responses to various parameters in the circuit, including magnet inductance and ability for time dependent resistance modeling of the load. The simulation was initially successfully verified with the conventional magnet with known inductance and resistance where discharged currents were limited to several kA.

### B. Training of Superconducting Magnets and QCD Baseline

Superconducting magnets still train [9], [10], [11] and this remains a major issue to resolve [12]. Since QCD is supposed to affect the training curve, a solid baseline for comparison is desired. A performance summary of magnet series tested at FNAL showed that "11 T" (dipoles) and "LARP" (quadrupoles) short models provided good reproducible training trends [10]. It was also concluded there that, to a good degree, coils inside magnets train independently, and coils in mirror magnets [13] train similarly to coils in "complete" (dipole/quadrupole) magnets. Thus, a mirror magnet is well suited for QCD testing. We chose to start with the "11 T" series as the training pattern baseline for QCD testing.

There are several features in magnet, or rather coil, training important in the current context, general discussion is found in [10]. When quench current is away from conductor limit coil training is largely independent on liquid helium temperature, i.e. quench current would be the same at 4.2 K and at 1.9 K. The behavior is drastically different close to the conductor limit - transitioning from higher to lower temperature after training, would initiate an additional (faster) training sequence. Damaged coils could exhibit variety of dependencies and features, depending on the nature of the damage, and current may be limited below conductor limit. However, all coils in the "11 T" baseline behaved "normally" in that respect with no abnormal dependencies observed.

### C. Superconducting Magnet Testing with and without QCD

A "mirror" magnet [13] from the "11 T" series [1, Chapter 8] was assembled specifically for QCD testing. It employed a coil which was fabricated as the last coil (#12) of the "11 T" program at FNAL many years ago and was never used. It was the third "11 T" mirror magnet assembled with similar parameters. The first low-voltage QCD discharges at up to few kA trip-current through this magnet were conducted on March 1$^{st}$, 2022, as part of magnet check out. The first discharge at spontaneous quench occurred on March 2$^{nd}$. All initial magnet training was at 4.5 K following the established baseline.

To compare to the baseline as directly as possible, QCD was discharged at quench detection time while ramping conditions (temperature, ramp rate) were kept nominal with respect to baseline. Simulations showed that the current boost needed 15-20 ms to reach its peak and that delaying magnet protection by 50 ms is safe for the magnet for quench currents below 12 kA. We did not have complete multi-physics simulation to know the expected effect of quench-back which was inevitable at such large current differential dI/dt. Thus, we did not know the expected resistance growth in the magnet, we conservatively ignored it while making protection assessments.

Before testing at high magnet current, we had to make major decisions based on partial or no information. Among those, we did not know the importance of the "over-current" (levels above the "quench current") shape or duration on performance/training. We hypothesized there may be some "effective" current, below the peak boost current, which represents the integral boost effect and is more relevant for training than the peak current; the only available reference [2] considered pulse time duration to be of importance. At this time, we had one magnet and one shot (test sequence) to investigate. Our main handle was settable QCD voltage, up to 1 kV, affecting the boost current. There was the possibility that even with high peak boost current we could be too low in "effective" current to observe any effect from QCD. On the other hand, the "effective" current may be close or equal to the peak current which may be high long enough to damage the magnet and halt any further QCD testing. There was also the remote possibility that the fast discharge process at high magnet current and QCD voltage may affect the magnet integrity negatively. In our steps we tried to navigate through those risks.

The first spontaneous quench with immediate QCD application did occur at expected current level (9 kA, [10]) and we chose QCD voltage of 800 V providing a substantial boost. Retrospectively, we found the resistance growth in the magnet to approximately follow linear trends: 0-35 mΩ from 5 to 22 ms after quench detection and 35-50 mΩ from 22 to 42 ms after detection; this dependence was embedded in the LTspice simulation along with negligibly small quench spot resistance. Fig. 1 then compares the updated simulation, with the observed real magnet current shape. The good description of current development gave us confidence to proceed with a higher boost current in the next quench allowing for longer "over-current" time. No abnormal behavior in monitored signals from the magnet was observed.

The second spontaneous quench occurred at current level well above the first one but the third went down, well below the expected level from the baseline. The QCD voltage was at the maximum 1 kV; no abnormal data signals were observed.

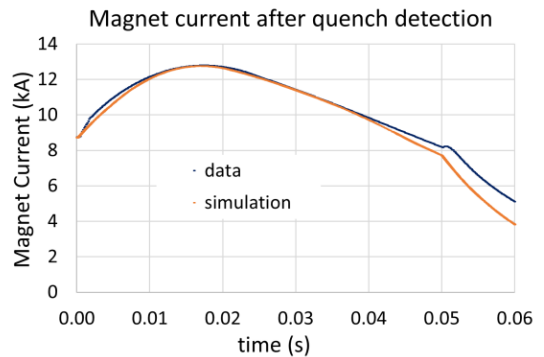

Fig. 1. Magnet current development in the first ramp, driven by QCD. There is 50 ms delay of dump resistor firing and 30 ms delay of protection heater firing but the latter has its own response time to affect the conductor. The simulation is not perfect: it assumes constant magnet inductance of 1.4 mH (measured 2.0/1.75 mH at ~ 4/7 kA), magnet resistance development is approximate.

After the third quench the QCD voltage was dropped to 500 V, able to boost the current to ~11 kA if no further magnet training. Several more quenches confirmed the magnet is at a current plateau, within a wide range. At this point our baseline approach was failing. We continued to follow our test plan and lowered the temperature to 1.9 K for further testing, initially keeping QCD voltage of 500 V. Then we moved on to perform several thermal cycles (TC).

Figure 2. shows the complete quench history of the mirror magnet at nominal ramp rate (20 A/s) and temperatures (4.5/1.9 K). The quenches at 1.9 K in the TC1 were all in a narrow current plateau at the same fraction of Short Sample Limit (SSL) as the 4.5 K level, namely ~70 %. We stopped using QCD in the last two 1.9 K quenches, the quench current levels remained the same. Consequent 4.5 K quenches returned to the current level observed earlier at 4.5 K. Quench current dependence on ramp rate was determined and was consistent with earlier 11 T magnet coils [1, Chapter 8], including mirror magnet coils. Conclusions at this point were that the coil reached conductor limit, albeit very low one, without training between 4.5 K and 1.9 K unlike other 11 T coils or any other accelerator magnet training ever observed. TC1 and all following thermal cycles ended at room temperature with the magnet remaining in the test facility cryostat.

TC2 training started at 1.9 K without any use of QCD. The magnet forgot its training practically entirely, which is unusual for $Nb_3Sn$ accelerator magnets, and needed 4 training quenches to reach the fraction of SSL observed in TC1. Quenches at 4.5 K re-confirmed conductor limitation as in TC1. QCD discharges were re-introduced for TC3 with capacitor voltage of 800 V and 500 V in the first and second quenches (1.9 K), respectively. The first quench of TC3 was at the same current level as in TC2. The second quench along with few more quenches were at conductor limit clearly indicating the effect of QCD on the training curve; later quenches at 4.5 K confirmed conductor limitation. The picture from TC1 to TC3 is quite unambiguous. In TC4 we did not use QCD and wanted to demonstrate again magnet training, but this time training was not fully forgotten by the magnet. Still, the first two quenches were identified as "training" based on quench location in the first layer. All limiting quenches in all TCs and at both temperatures were identified in the outer coil layer, including the last ones at 4.5 K in TC4.

### III. QCD DISCUSSIONS

#### A. Discussion on Magnet and QCD Performance

The mirror magnet coil tested clearly underperformed compared to other "11 T" coils (as presented in [10]). A question arises if this has to do with QCD in any way.

The QCD capacitor discharge in the second quench (highest boost current reached) drove the current rise at dI/dt ~ 1 MA/s in the first two ms and ~0.5 MA/s in the next 2 ms, easing substantially after that. The average increase to peak was 0.3 MA/s. In comparison, CLIQ discharges, which have somewhat similar dI/dt characteristics in the first 10-15 ms have an average increase to peak of 0.15 MA/s (dependencies exist, data from non-"11 T" magnets). Moreover, regular quench protection itself in small magnets drives dI/dt as ~ 0.5 MA/s in the first 5 ms. All this is to say QCD pushes to higher differential current increases but those are still of the same order as known applications. Analysis of energy flow and energy density in the magnet, including the QCD energy introduced to the system, shows that the magnet bulk temperature never exceeded 150 K and the hot-spot temperature was below 210 K after current dump. We used the cable enthalpy estimates from ([14], Fig 13) and quench integral calculations from [15].

The coil used in the present test featured all improvements made during the "11 T" program. However, it was fabricated by a partially different (new) team at the time. The team assembling the magnet was also different than earlier magnets. Fig. 3 shows a prominent non-planarity feature of the coil and uncharacteristic cracks observed in the non-lead end, outer layer of the coil. This area is consistent with all limiting quench locations as observed by quench antenna [16]. Quenching at this area yielded a characteristic QA signal development pattern, up to quench detection, in several channels. The same pattern was identified immediately after the very first quench in the inner layer too, pointing to pre-existing conditions for the under-performance. We hypothesize that the coil shimming corrections with Kapton layers, based on average deviations, could not fix in full the abnormal divergence from flatness observed on the non-lead end and this caused tension and over-stress on the coil non-lead end outer layer.

QCD developments aimed to investigate timing characteristics of accelerator magnet training. The typical magnet ramp rate range is 1-300 A/s (nominally ~ 10 A/s), and data suggest magnets train regardless of ramp rate. Average current gains after spontaneous training quenches vary but they are of the order of 100 A. Thus, training mechanisms act, nominally, within 10 s. With QCD one can test characteristic times up to tens of ms with low limit driven by practical limitations. QCD does not change current distribution across the magnet, and truly emulates known training conditions but at higher ramp rate. QCD results so far show coil training can be affected at ~30 ms timescale. If it holds that coils train independently [10] then we can conclude that CLIQ [7] too affects training but training quenches in magnets shall occur predominantly in coils with no over-current (due to CLIQ).

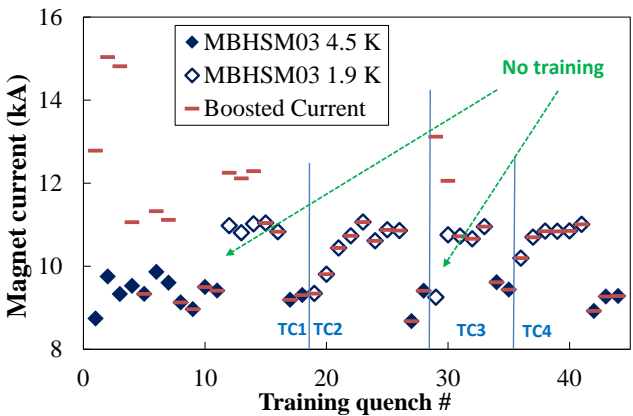

Fig. 2. Spontaneous quenches at nominal ramp rate – magnet current at quench detection vs quench number; boosted current at its peak is shown as well; the two currents differ only if QCD is applied. All four thermal cycles (TC) are included in the plot. Lack of training quenches is indicated.

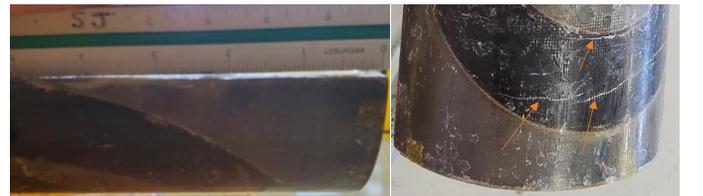

Fig. 3. Unusual non-conformities on the coil. Left: significant non-planarity at the coil mid-plane (non-lead end), since coil fabrication; Right: cracks (pointed to by orange arrows) on the coil non-lead end, outer layer, post-testing.

Another effect of QCD in support of training reduction can be seen through the Keiser effect [17], [18] which is well established in magnets – they become "quieter" at "known" current levels. Fig. 4 presents acoustic (mechanical) activity recorded during current ramps in TC1 and TC3 (with QCD) and TC2 (no QCD). After QCD is applied, the magnet is quieter in ramps above previously reached current levels.

### B. Discussion on Future of QCD

There are more tests planned to perform with QCD on small magnet models: we still do not know what the relevance of over-current length or shape on training is; nor we know how different magnet designs may behave. However, another important question is on applications to large accelerator magnets. The main hurdle is their larger inductance which limits the current boost level along with the induced by QCD large normal zone. Although the existing QCD can be upgraded in terms of capacitance (C), voltage (U) increase would be more relevant as an upgrade (energy ~ $CU^2$). Fig. 5 shows simulations of QCD with nominal and upgraded parameters for a magnet of 35 mH (a typical HL-LHC $Nb_3Sn$ quadrupole). There is a non-negligible positive effect that can be expected from a higher voltage QCD likely at the expense of magnet insulation requirements. Novel designs with decreased inductance, like bi-filar windings [19], or multi-magnet circuit designs could benefit more significantly by QCD as it is.

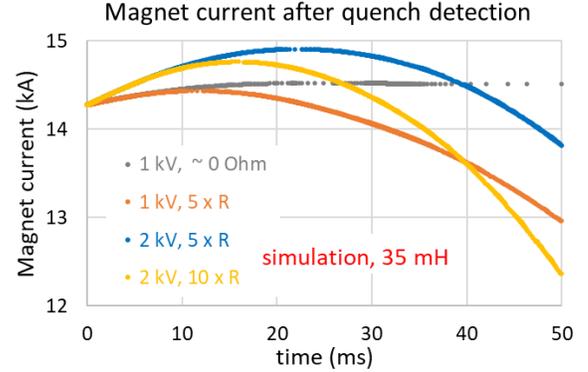

Fig. 5. Simulations with QCD on a 35 mH load at 14 275 kA "first" quench current (both targeting a HL-LHC quadrupole magnet). "R" is the resistance dependence described earlier; actual resistance growth is not known. Compared to the discussed mirror magnet, HL-LHC magnets have ~15 times the resistance at 300 K, and about twice as large RRR. With 2 kV discharge the current is boosted over 500-600 A with 30-40 ms over-current duration.

Although QCD is in effect also a protection device its use does require additional delay for other protection mechanisms. Depending on the concrete case this may not be a real problem – heater protection, for instance, has internal delays. More importantly, QCD does not need a quench to operate. In this mode, there is no hot-spot, per se, and there is no quench detection delay. Operating QCD with series of step-like high-current trips to eventually reach a quench plateau in a magnet will be tested in following magnet experiments. Ultimately, protection issues do not appear to be prohibiting for using QCD though deeper case-by-case analysis is needed.

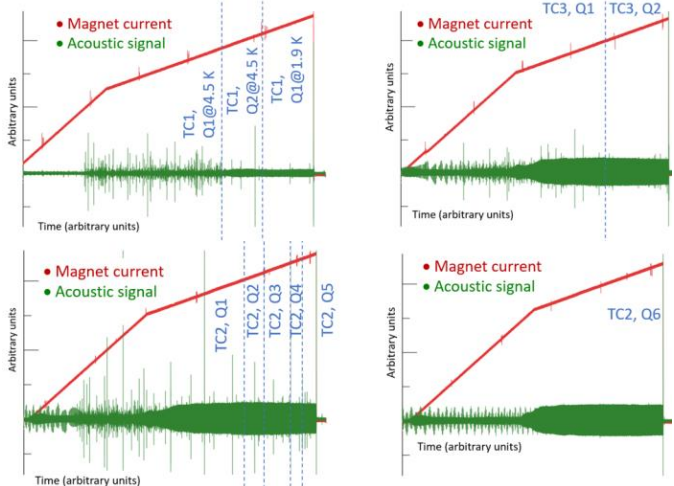

Fig. 4. Acoustics data from the three consecutive thermal cycles. The plots are "stitched" together (note dotted lines) from different ramps to quench ("Qi") at the previously highest quench current for that TC. The highest reached currents on all figures are at approximately the same level (quench plateau). Once QCD is applied (TC1 and TC3) the magnet becomes quieter. The lower right figure shows data from the whole ramp to quench after the magnet reached quench current plateau in TC2. The instrumental noise level was not well controlled over time for both current and acoustics.

### IV. CONCLUSIONS

A new device (QCD) aiming to affect superconducting magnet training, has been commissioned, and tested. Results support the notion that QCD-like discharges could eliminate training in superconducting magnets. Possible negative effects of QCD on magnets were investigated but no clear evidence or clues were found. More experiments are needed to determine the limits of capacitor discharges to affect training. Larger accelerator magnets can also benefit from QCD, but they probably need to have better insulation scheme allowing for higher QCD voltage. Novel magnet designs with lower inductance would be more susceptible to QCD.

### ACKNOWLEDGMENTS

We thank Howie Pfeffer for actively supporting the phase of QCD commissioning, Matt Kufer for invaluable electrical engineering and help with LTSpice simulations, Chris Jensen for leading the overall efforts; MSD and T&I colleagues who contributed to QCD commissioning, data taking and engaged in analysis discussions; LBNL for help with instrumentation.